\documentclass[a4paper,11pt]{article}
\usepackage{pos}
\usepackage{graphicx,amsmath,amsfonts}
\usepackage{slashed}

\allowdisplaybreaks[2]
\usepackage{calrsfs}
\DeclareMathAlphabet{\pazocal}{OMS}{zplm}{m}{n}
\def\pt{\widetilde p}
\def\pa{p_A}
\def\pn{p_N}

\title{Nuclear effects on longitudinal-transverse structure function ratio
in the deuteron}

\author[a,b,c]{S. Kumano}
\affiliation[a]{Quark Matter Research Center,
    Institute of Modern Physics, Chinese Academy of Sciences,\\
    Lanzhou, 730000, China}
\affiliation[b]{ Southern Center for Nuclear Science Theory,
    Institute of Modern Physics, Chinese Academy of Sciences,\\
    Huizhou, 516000, China}
\affiliation[c]{KEK Theory Center, Institute of Particle and Nuclear Studies, KEK,
    Oho 1-1, Tsukuba, 305-0801, Japan}
\emailAdd{kumanos@impcas.ac.cn}

\abstract{
Nuclear modifications of the nucleon's structure function $F_2^N$
have been investigated mainly since the discovery of the EMC nuclear effect
in 1983, and there were many experimental measurements from the deuteron to 
a heavy nucleus. Now, the details of the modifications of $F_2^N$ are known
from small $x$ to large $x$. On the other hand, it is taken as granted
that a nuclear modification does not exist for 
the longitudinal-transverse structure function ratio $R_N=F_L^N/(2xF_1^N)$. 
However, such a nuclear modification does exist theoretically.
A nucleon in a nucleus moves in any space direction,
which is not necessary the longitudinal direction 
along the virtual-photon momentum in charged-lepton scattering.
Because of this transverse Fermi motion, the longitudinal
and transverse structure functions mix with the mixture probability 
proportional to the nucleon's transverse-momentum squared 
$\vec p_T^{\,\, 2}/Q^2$.
In this paper, the nuclear modifications are shown numerically
for the deuteron by using a standard convolution description.
The magnitude of the modifications is of the order of a few percent
in the deuteron; however, they should be large in large nuclei.
In handling high-energy nuclear data, 
such nuclear modifications need to be taken into account
for a precise determination of physical quantities.
Now, the longitudinal-transverse structure-function ratio
and tensor-polarized experiments are under preparation 
for the deuteron at JLab. We hope that such effects will be 
confirmed experimentally for not only for the deuteron
but also for larger nuclei.
}

\FullConference{
The 26th international symposium on spin physics (SPIN2025)\\
21--26 September, 2025\\
Qingdao (Tsingtao), Shandong Province, China\\}

\begin{document}
\maketitle

\section{Introduction}
\label{introduction}

Until the end of 1970's, nuclear modifications of structure functions
were considered negligible because the average binding energies
and Fermi momenta are small in comparison with
a typical energy scale of deep inelastic scattering (DIS).
Therefore, it was rather surprising that the European Muon Collaboration (EMC)
found significant nuclear effects in the structure function $F_2^N$,
where $N$ indicates the nucleon, in 1983. 
Such a nuclear modification is now called the nuclear EMC effect.
Because it was considered to be a big discovery as
the first quark signature in nuclear phenomena, 
there were a flood of papers on the effect just after the EMC fining. 
However, a conservative theoretical interpretation 
in terms of nuclear binding and Fermi motion was proposed 
a few years later, and they could be considered as the major sources of 
the nuclear modifications at medium and large $x$,
where $x$ is the Bjorken scaling variable.
There should be also 
internal nucleon modification effects within the nuclear medium,
but it is not straightforward to separate them from
the binding and Fermi-motion effects experimentally.
Now, there are many experimental data on $F_2^A$,
where $A$ indicates a nucleus, from small $x$ to large $x$ 
and from the deuteron to a large nucleus. 
There are various theoretical works
to interpret the nuclear modifications of $F_2^N$.

The structure function $F_2^{N,A}$
contains both longitudinal and transverse components.
For the longitudinal-transverse structure-function ratio 
$R_N=F_L^N/(2xF_1^N)$, one usually assumes
that there is no nuclear modification even now.
In the past, this issue was discussed extensively
around the year of 2000. However, there were no nuclear effect 
in the charged-lepton data in 2003 \cite{HERMES:1999bwb},
the neutrino DIS data in 2001 \cite{CCFRNuTeV:2001njk},
and the JLab (Thomas Jefferson National Accelerator Facility)
analysis in 2007 \cite{Tvaskis:2006tv}.
On the other hand, there was a theoretical proposal
that nuclear modifications should exist in 2003
\cite{Ericson:2002ep} despite the negative experimental results.
The reason is the following. The longitudinal and transverse
structure functions for the nucleon are defined by taking
the virtual-photon momentum direction along the $z$ axis
with the nucleon at rest or in the center-of-momentum frame.
For a nuclear target, they are defined in the same way.
However, a nucleon in the nucleus moves in any space direction,
so that the longitudinal and transverse structure functions
mix in the nucleus. Because this mechanism is associated
with the transverse motion of the nucleon, such effects 
are expected to be proportional to $\vec p_{T}^{\, 2}/Q^2$.

It is worth to bring this nuclear-modification topic forward now
\cite{Kumano:2025qzm}.
In analyzing nuclear data, the nucleon's longtidutinal-transverse 
ratio $R_N$ is used even for heavy nuclei by assuming that 
there is no nuclear modification.
However, it is not appropriate for a precise determination
of physical quantities from nuclear data
because the nuclear modification certainly exists theoretically
\cite{Ericson:2002ep,Kumano:2025qzm}.
Because the ratio will be measured for the deuteron
at JLab in the near future \cite{R-Jlab-pro-R} and the polarized deuteron
experiment is under preparation \cite{Poudel:2025nof}, I show
such nuclear modifications numerically for the deuteron in this work.
The theoretical formalism is explained in Sec.\,\ref{formalism}, 
numerical results are shown in Sec.\,\ref{results},
and they are summarized in Sec.\,\ref{summary}.

\section{Formalism}
\label{formalism}

The lepton-hadron DIS cross section is described by the hadron tensor
$W^{A,N}_{\mu\nu} (p_{_{A,N}}, q)$ where $p_{_{A,N}}$ is the nuclear
or nucleon's momentum and $q$ is the momentum transfer.
The hadron tensor in the charged-lepton DIS is decomposed into
two structure functions $W^{A,N}_1$ and $W^{A,N}_2$ as
\begin{align}
W^{A,N}_{\mu\nu} (p_{_{A,N}},  q)  =
  - W^{A,N}_1 (p_{_{A,N}}, q) 
  \left ( g_{\mu\nu} - \frac{q_\mu q_\nu}{q^2} \right )
+ W^{A,N}_2 (p_{_{A,N}}, q) \, \frac{\pt_{_{A,N} \mu} 
  \, \pt_{_{A,N} \nu}}{p_{_{A,N}}^2} 
\ ,
\label{eqn:hadron}
\end{align}
where $\pt_{\mu} = p_{\mu} -(p \cdot q) \, q_\mu /q^2$.
The hadron tensor with the photon helicity 
$\lambda$ is defined by multiplying the photon polarization vector 
$\varepsilon_\lambda^{\mu}$ as
\begin{align}
W^{A,N}_\lambda (p_{_{A,N}}, q) 
            = \varepsilon_\lambda^{\mu *} \varepsilon_\lambda^{\nu} 
                  W^{A,N}_{\mu\nu} (p_{_{A,N}}, q) .
\label{eqn:W_lambda}
\end{align}
The transverse and longitudinal structure functions 
$W^{A,N}_T$ and $W^{A,N}_L$
are defined by these helicity functions
and also by $W^{A,N}_{1}$, and $W^{A,N}_{2}$ as
\begin{align}
W^{A,N}_T = 
    \frac{ W^{A,N}_{\lambda=+1} + W^{A,N}_{\lambda=-1} }{2}
     = W^{A,N}_{1} ,
\ \ \ 
W^{A,N}_L = W^{A,N}_{\lambda=0}
     = \left ( 1 + \frac{\nu_{_{A,N}}^2}{Q^2} \right) W^{A,N}_2 - W^{A,N}_1 .
\end{align}
The variables $\nu_N$ and $\nu_A$ are given by
$ \nu_A = p_{_A} \cdot q / \sqrt{p_{_A}^{\,2}} = \nu$ and
$ \nu_N = p_{_N} \cdot q / \sqrt{p_{_N}^{\,2}}$ .
In these days, structure functions $F_{1,2,T,L}$ are usually used instead of
$W_{1,2,T,L}$, and they are related as
\begin{align}
F_1^{A,N} & = F_T^{A,N}  = \sqrt{p_{_{A,N}}^2} \, W_1^{A,N} ,
\ \ 
F_2^{A,N}  = \frac{p_{_{A,N}} \cdot q}{\sqrt{p_{_{A,N}}^2}} \, W_2^{A,N} ,
\nonumber \\
F_L^{A,N} & = \left ( 1 + \frac{Q^2}{\nu_{_{A,N}}^2} \right ) F_2^{A,N} 
            - 2 x_{_{A,N}} F_1^{A,N} .
\end{align}
Then, the longitudinal-transverse structure function ratio is
calculated as
\begin{align}
R_{A,N} (x_A, Q^2) = \frac{F_L^{A,N} (x_{A,N}, Q^2)}
 {2 \, x_{A,N} F_1^{A,N} (x_{A,N}, Q^2)}  .
\label{eqn:RAN}
\end{align}
The used kinematical variables are $x_A$, $x_N$, $x$, and $y$ defined by
\begin{align}
x_A  = \frac{Q^2}{ 2 \, p_A \cdot q}, \ \ \ 
x_N  = \frac{Q^2}{ 2 \, p_N \cdot q} = \frac{x}{y}, \ \ \ 
x    = \frac{Q^2}{2 \, M_N \nu},  \ \ \ 
y    = \frac{p_N \cdot q}{M_N \, \nu},
\end{align}
where $M_N$ and $M_A$ are nucleon and nuclear masses, respectively.

A standard method for describing the nuclear structure functions 
is to use the convolution formalism with the spectral function
$S(\pn)$ as
\begin{align}
W^A_{\mu\nu} (\pa, q) 
  = {\displaystyle\int} d^4 \pn \, S(\pn) \, W^N_{\mu\nu} (\pn, q).
\end{align}
The technical details of this convolution formalism are explained 
in Refs.\,\cite{Hirai:2010xs,Cosyn:2017fbo}.
A simple spectral function is given by the momentum-space wave function
$\phi_i$ as
\begin{align}
S (p_N)  = \sum_i | \phi _i (\vec p_N) |^2  \delta 
   \left ( p_N^{\, 0} - M_A + \sqrt{M_{A-i}^{\ 2} 
          +\vec p_N^{\ 2}} \, \right ).
\label{eqn:spectral}
\end{align}
The summation is taken over the nucleon $i$ in the nucleus,
and the mass factor $M_{A-i}$ is given by 
the separation energy $\varepsilon_i$ 
by $\varepsilon_i = (M_{A-i}+M_N) - M_A$.
In the following numerical analysis, 
the average separation energy is taken 
$\varepsilon_i \rightarrow \left< \varepsilon \right>$
and the non-relativistic kinematic is assumed
for the squared root in Eq.\,(\ref{eqn:spectral}).
Then, $p_N^0$ and $\left< \varepsilon \right>$ are related by
$ p_N^{\, 0} = M_N - \left< \varepsilon \right> 
   - \vec p_N^{\ 2} / (2 M_{A-1})$ with the replacement
$M_{A-i} \to M_{A-1}$.
Next, multiplying the projection operators of $W_{1,2}^A$
and converting them to $F_{1,2,L}$, we have the nuclear
structure functions expressed by the nucleon ones as
\begin{align}
\left(
    \begin{aligned}
      \,      F_2^A(x_A,Q^2) \, \\
      \, 2x_A F_1^A(x_A,Q^2) \, \\
      \,      F_L^A(x_A,Q^2) \,
    \end{aligned}
\right)
& \, = \int_x^A dy \,
\left(
    \begin{aligned}
      \, f_{22} (y) \, & \ \ \ \, 0 \,      & \, 0 \ \ \ \,  \\
      \, 0 \ \ \ \     & \, f_{11} (y) \, & \, f_{1L} (y) \,\\
      \, 0 \ \ \ \    & \, f_{L1} (y) \, & \, f_{LL} (y) \,
    \end{aligned}
\right) \,
\left(
    \begin{aligned}
      \,               F_2^N(x/y,Q^2) \, \\
      \, 2 \, \frac{x}{y} F_1^N(x/y,Q^2) \, \\
      \,               F_L^N(x/y,Q^2) \,
    \end{aligned}
\right) .
\label{eqn:12L-convolution}
\end{align}
The nucleon's lightcone momentum distributions
$f_{22}$, $f_{LL}$, $f_{11}$, $f_{L1}$, and $f_{1L}$
in the nucleus are given
\begin{align}
f_{22} (y) & = \int_0^\infty dp_{N\perp} 2 \pi p_{N\perp} y
             \frac{M_N \nu}{|\vec q \, |} 
             \left | \phi (\vec p_N) \right |^2
  \left [ \frac{2 \, (x/y)^2 \, \vec p_{N\perp}^{\ 2}}{(1+Q^2/\nu^2)Q^2}
 + \left ( 1 + \frac{2 \, (x/y) \, 
        p_{N\parallel}}{\sqrt{Q^2+\nu^2}} \right )^2 \right ],
\nonumber \\
f_{LL} (y) & = f_{11} (y) 
    = \int_0^\infty dp_{N\perp} 2 \pi p_{N\perp} y
             \frac{M_N \nu}{|\vec q \, |} 
             \left | \phi (\vec p_N) \right |^2
    \left ( 1 + \frac{\vec p_{N\perp}^{\ 2}}{\pt_N^{\,\, 2}} \right ) ,
\nonumber \\
f_{L1} (y) & = f_{1L} (y)
    = \int_0^\infty dp_{N\perp} 2 \pi p_{N\perp} y
             \frac{M_N \nu}{|\vec q \, |} 
             \left | \phi (\vec p_N) \right |^2
             \frac{\vec p_{N\perp}^{\ 2}}{\pt_N^{\,\, 2}} .
\label{eqn:fL1}
\end{align}
The momentum $\pt_{\mu}$ squared is given by $Q^2$ as 
\begin{align}
\pt_N^{\,\, 2} = \frac{Q^2}{4 x_N^2} 
                 \left( 1+\frac{4 x_N^2 p_N^2}{Q^2} \right)
            \simeq \frac{Q^2}{4 x_N^2} 
                 \left( 1+\frac{4 x_N^2 M_N^2}{Q^2} \right) .
\label{eqn:ptN2-Q2}
\end{align}
The longitudinal and transverse structure functions mix
with each other, and the mixture coefficient is proportional to 
$\vec p_{N\perp}^{\ 2} / \pt_N^{\,\, 2} 
 \sim \vec p_{N\perp}^{\ 2} / Q^2$.
Therefore, the structure-function mixture between
$F_1^N$ and $F_L^n$ in Eq.\,(\ref{eqn:12L-convolution}) 
does not occur in the scaling limit $Q^2 \to \infty$.
Now, the theoretical tool is ready to calculate
the nuclear structure functions by Eq.\,(\ref{eqn:12L-convolution}) and 
the longitudinal-transverse ratio $R_A$ by Eq.\,(\ref{eqn:RAN}).

\section{Results}
\label{results}

\begin{figure}[t]
\begin{minipage}[c]{0.45\textwidth}
     \hspace{-0.20cm}
     \includegraphics[width=7.0cm]{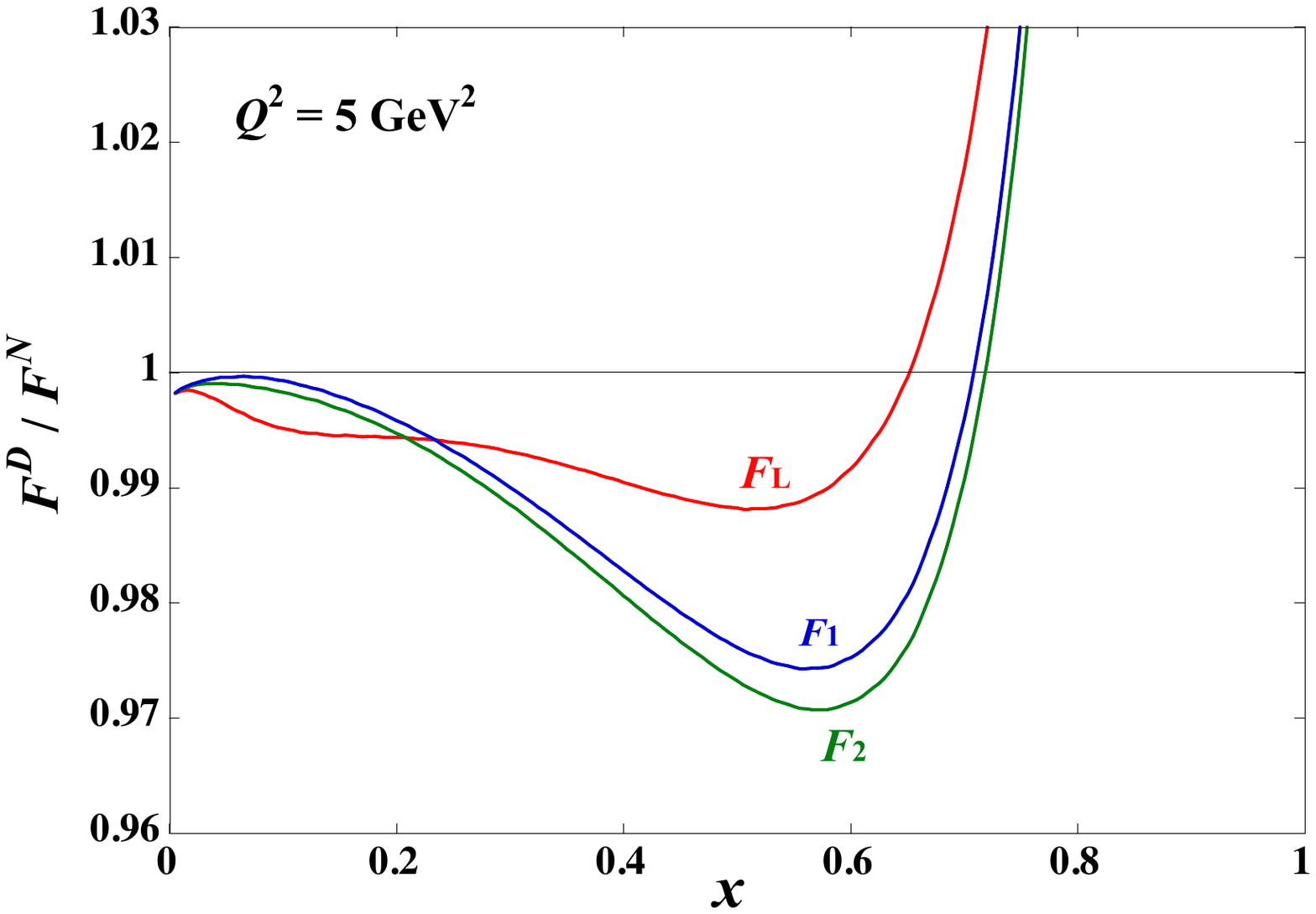}
\vspace{-0.90cm}
\caption{Nuclear modifications of 
$F_1^N$, $F_2^N$, and $F_L^N$ for the deuteron at $Q^2=5$ GeV$^2$.}
\label{fig:modifications-F12L}
\vspace{-0.4cm}
\end{minipage}
\hspace{0.50cm}
\begin{minipage}[c]{0.45\textwidth}
     \hspace{-0.20cm} \vspace{-0.05cm}
     \includegraphics[width=7.0cm]{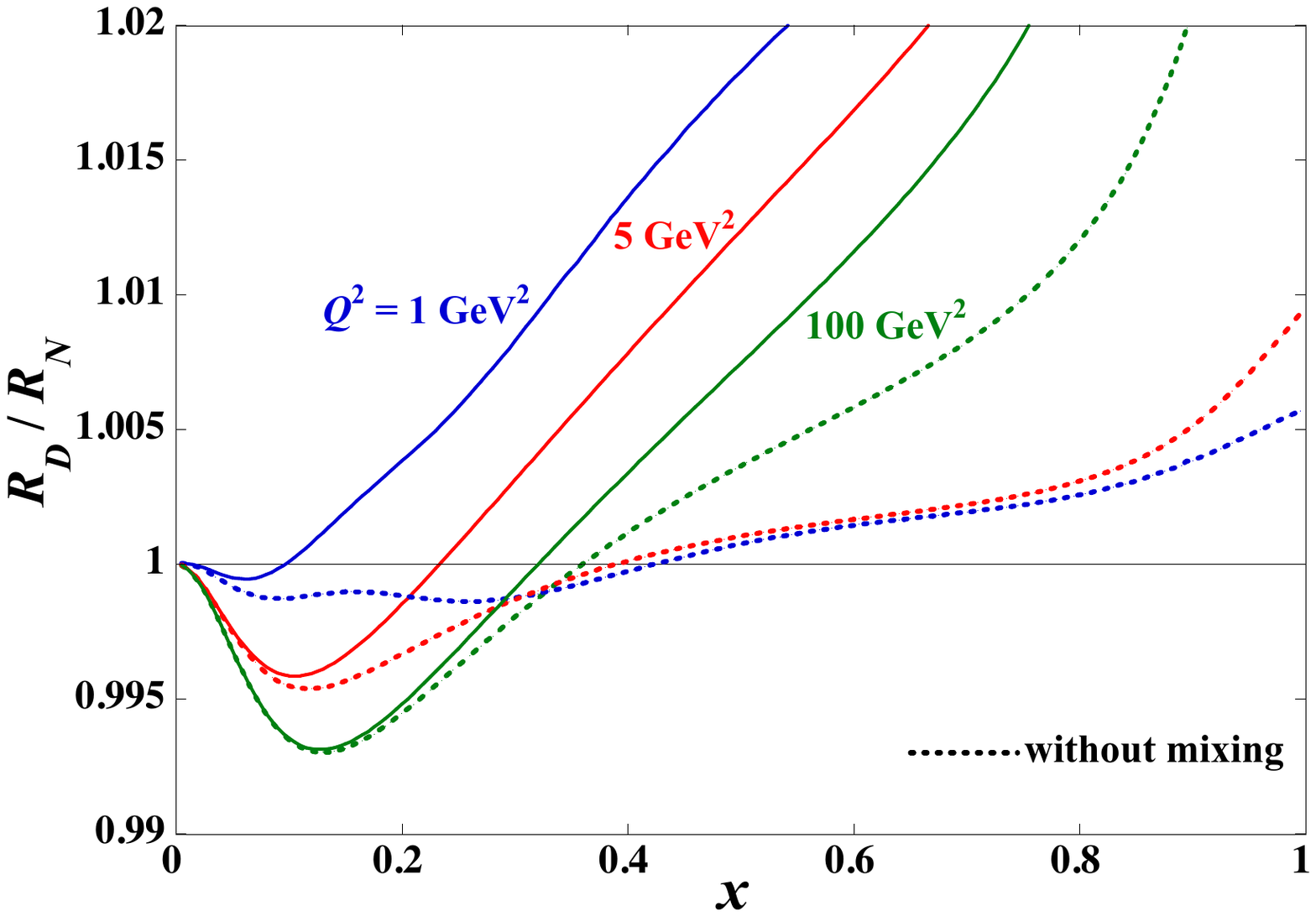}
\vspace{-0.90cm}
\caption{Nuclear modifications of 
$R_N$ for the deuteron 
at $Q^2=1$, 10, and 100 GeV$^2$.}
\label{fig:modifications-R-1-10-100}
\vspace{-0.4cm}
\end{minipage}
\end{figure}

In order to calculate the nuclear modifications for the deuteron,
it is necessary to supply the deuteron wave functions $\phi_i$
and the structure function $F_1^N$, which is given by
$F_1^N (x_N,Q^2) =  ( 1+Q^2/\nu_N^2 ) \, 
 F_2^N (x_N,Q^2) / [2 \, x_N \, \{ 1+R_N (x_N,Q^2) \} ]$.
In the leading order of the QCD running coupling constant $\alpha_s$,
$F_2^N$ is calculated by using unpolarized PDFs of MSTW08.
As for $R_N$, the SLAC parametrization of 1990 is used.
As for the deuteron's $\phi_i$, the Bonn wave function is used.
Calculating Eqs.\,(\ref{eqn:12L-convolution}) and (\ref{eqn:fL1}),
we obtain the nuclear modifications in 
Fig.\,\ref{fig:modifications-F12L} at $Q^2 = 5$ GeV$^2$.
The nuclear modifications due to the nuclear binding and Fermi motion,
which are contained within the spectral function of 
Eq.\,(\ref{eqn:spectral}), are well investigated in $F_2^A$. 
The negative modifications at medium $x$ ($\sim 0.4$) come mainly
from the nuclear binding, and the steep positive rises
at large $x$ ($\sim 0.7$) are caused by the nucleon's Fermi motion
in the nucleus.
The structure function $F_2^A$ has both longitudinal and transverse
components. The nuclear modifications of the transverse component
$F_1^N$ are similar to the one of $F_2^N$ 
in Fig.\,\ref{fig:modifications-F12L}. 
However, the modifications of the longitudinal component $F_L^N$ 
are very different from the ones of $F_2^N$. 

Next, nuclear modifications of the longitudinal-transverse
structure-function ratio $R_N$ are calculated by using
Eq.\,(\ref{eqn:RAN}), and the results are shown in 
Fig.\,\ref{fig:modifications-R-1-10-100}
at $Q^2=1$, 10, and 100 GeV$^2$.
The dashed curves are obtained by terminating the mixing terms,
namely by taking $\vec p_{N\perp}^{\ 2} / \pt_N^{\,\, 2} \to 0$
in $f_{LL} (y)$, $f_{11} (y)$, $f_{L1} (y)$, and $f_{1L} (y)$,
and by taking $[ \cdot\cdot\cdot ] \to 1$ in $f_{2} (y)$.
The differences between the solid and dashed curves become
smaller as $Q^2$ increases because it comes from 
the longitudinal-transverse admixture proportional to
$\vec p_{N\perp}^{\ 2} / Q^2$. 
However, the modifications themselves are not small even 
at $Q^2=100$ GeV$^2$. 
Although the nucleon momentum distributions are similar
in both longitudinal and transverse cases
in Eq.\,(\ref{eqn:fL1}), 
the $x$-dependent functions are very different
between the longitudinal and transverse structure functions,
which leads to the modifications even without the mixture
as shown by the dashed curves at $Q^2=100$ GeV$^2$.
From these results, we find that there are two major factors 
in the nuclear modifications 
of $R_N$ in the convolution formalism:
\vspace{-0.20cm}
\begin{itemize}
\setlength{\itemsep}{-0.10cm}
\item[(1)] One is the admixture of the longitudinal and
transverse structure functions of the nucleon. 
\item[(2)] The other is that the $x$-dependent functional forms
are different between the longitudinal and
transverse structure functions, which results in the differences
after the convolution integral.
\vspace{-0.20cm}
\end{itemize}
In this work, I did not discuss the small-$x$ region,
where a different physics mechanism contributes,
and my studies are focused on the medium- and large-$x$ regions,
where the convolution model works.
In general, the nuclear modifications are large at small $Q^2$
and a typical effect is of the order of a few percent in the deuteron,
although it depends much on $x$. As $x$ becomes large, 
the nuclear effect becomes large. In general, it is not a large effect 
in the deuteron in Fig.\,\ref{fig:modifications-R-1-10-100}.
However, it should become larger as the nucleus 
becomes larger. In addition, there are higher-twist effects 
and target-mass corrections at large $x$ \cite{Kumano:2025qzm}, 
so that these effects should be considered as well as 
the nuclear corrections of $R_N$.

Although it has been assumed that the nuclear modification 
does not exist for the longitudinal-transverse structure-function
ratio $R_N$, it certainly exists as shown in this work.
In analyzing nuclear cross sections, one needs to be careful
to take into account such modifications.
The deuteron experiments are under preparation at JLab,
one needs to be careful about such nuclear effects.
In general, if the target is a heavy nucleus, one should
consider them in analyzing cross sections.

\vfill\eject
\section{Summary}
\label{summary}

It was shown numerically that the nuclear modifications should exist 
in the deuteron for the longitudinal-transverse structure-function 
ratio $R_N$ by using the standard convolution model. 
In particular, the transverse motion of a nucleon in a nucleus
gives rise to the mixture of the longitudinal and transverse 
structure functions with the probability proportional to 
$\vec p_{N\perp}^{\ 2} / Q^2$, which is one of sources of
the nuclear modifications of $R_N$. Another source is 
the difference of the $x$-dependent functional forms
between the longitudinal and transverse structure functions 
in the convolution integral.
The nuclear modifications are large at small $Q^2$, of the order of
a few GeV$^2$, and at large $x$. The modifications are typically
a few percent in the deuteron, but they should be larger 
as the nucleus becomes larger. In analyzing nuclear cross sections,
one need to be careful about such nuclear effects
for a precise determination of physical quantities.
We also hope that the nuclear modification of $R_N$ will
be found experimentally.




\begin{thebibliography}{99}
\setlength{\itemsep}{-0.01cm}
\setlength\baselineskip{14pt}
\vspace{-0.22cm}
\bibitem{HERMES:1999bwb}
    K. Ackerstaff et al. (HERMES Collaboration),
    \href{https://doi.org/10.1016/S0370-2693(99)01493-8}
    {\emph{Phys. Lett. B} {\bf 475} (2000) 386};
     A. Airapetian et al.,
    \href{https://doi.org/10.1016/j.physletb.2003.06.044}
    {\emph{Erratum: Phys. Lett. B} {\bf 567} (2003) 339}.
\bibitem{CCFRNuTeV:2001njk}
    U. K. Yang et al. (CCFR/NuTeV Collaboration),
    \href{https://doi.org/10.1103/PhysRevLett.87.251802}
    {\emph{Phys. Rev. Lett.} {\bf 87} (2001) 251802}.
\bibitem{Tvaskis:2006tv}
    V. Tvaskis et al.,
    \href{https://doi.org/10.1103/PhysRevLett.98.142301}
    {\emph{Phys. Rev. Lett.} {\bf 98} (2007) 2003}.
\bibitem{Ericson:2002ep}
    M. Ericson and S. Kumano,
    \href{https://doi.org/10.1103/PhysRevC.67.022201}
    {\emph{Phys. Rev. C} {\bf 67} (2003) 022201}.
\bibitem{Kumano:2025qzm}
    S. Kumano,
    \href{https://doi.org/10.1103/xgbh-grqx}
    {\emph{Phys. Rev. C} {\bf 113} (2026) 015206}.
\bibitem{R-Jlab-pro-R} 
     R. Ent {\it et al.}, 
     \href{https://www.jlab.org/exp_prog/proposals/proposal_updates/PR12-06-104_pac36.pdf}
     {Update of JLab proposal E12-06-104 (2010)}.
\bibitem{Poudel:2025nof}
    J. Poudel, A. Bacchetta, J.-P. Chen, 
    and N. Santiesteban, 
    \href{https://doi.org/10.1140/epja/s10050-025-01558-w}
    {\emph{Euro. Phys. J. A} {\bf 61} (2025) 81}.
\bibitem{Hirai:2010xs}
    M. Hirai, S. Kumano, K. Saito, and T. Watanabe,
    \href{https://doi.org/10.1103/PhysRevC.83.035202}
    {\emph{Phys. Rev. C} {\bf 83} (2011) 035202}.
\bibitem{Cosyn:2017fbo}
    W. Cosyn, Yu-Bing Dong, S. Kumano, and M. Sargsian,
       \href{https://doi.org/10.1103/PhysRevD.95.074036}
       {\emph{Phys. Rev.} {\bf D 95} (2017) 074036}. 
\end{thebibliography}
\end{document}